\documentclass{emulateapj}

\newcommand{\name}{SDSSJ0221}

\begin{document}
\title{Characterizing a Dramatic $\Delta V\sim-9$ Flare on an Ultracool Dwarf Found by the ASAS-SN Survey$^{\dagger}$}
\shorttitle{A Dramatic Flare on an Ultracool Dwarf}

\author{Sarah J. Schmidt\altaffilmark{*,1,3}, Jose L. Prieto\altaffilmark{2,4}, K. Z. Stanek\altaffilmark{1,5}, Benjamin J. Shappee\altaffilmark{1,6}, Nidia Morrell\altaffilmark{7}, Daniella C. Bardalez Gagliuffi\altaffilmark{3,8}, 
{C.~S.~Kochanek}\altaffilmark{1,5},
{J.~Jencson}\altaffilmark{1},
{T.~W-S.~Holoien}\altaffilmark{1},
{U.~Basu}\altaffilmark{1,9},
{John.~F.~Beacom}\altaffilmark{1,5,10},
{D.~M.~Szczygie{\l}}\altaffilmark{11},
{G.~Pojmanski}\altaffilmark{11},
{J.~Brimacombe}\altaffilmark{12},
{M.~Dubberley}\altaffilmark{13},
{M.~Elphick}\altaffilmark{13},
{S.~Foale}\altaffilmark{13},
{E.~Hawkins}\altaffilmark{13},
{D.~Mullins}\altaffilmark{13},
{W.~Rosing}\altaffilmark{13},
{R.~Ross}\altaffilmark{13},
{Z.~Walker}\altaffilmark{13}
}

\altaffiltext{*} {schmidt@astronomy.ohio-state.edu}
\altaffiltext{1}{Department of Astronomy, Ohio State University, 140 West 18th Avenue, Columbus, OH 43210}
\altaffiltext{2}{Department of Astrophysical Sciences, Princeton University, 4 Ivy Lane, Peyton Hall, Princeton, NJ 08544}
\altaffiltext{3}{Visiting Astronomer at the Infrared Telescope Facility, which is operated by the University of Hawaii under Cooperative Agreement no. NNX-08AE38A with the National Aeronautics and Space Administration, Science Mission Directorate, Planetary Astronomy Program.}
\altaffiltext{4}{Carnegie-Princeton Fellow}
\altaffiltext{5}{Center for Cosmology and AstroParticle Physics, The Ohio State University, 191 W.\ Woodruff Ave., Columbus, OH 43210, USA}
\altaffiltext{6}{NSF Graduate Fellow}
\altaffiltext{7}{Las Campanas Observatory, Carnegie Observatories, Casilla 601, La Serena, Chile}
\altaffiltext{8}{Center for Astrophysics and Space Science, University of California San Diego, La Jolla, CA 92093, USA }
\altaffiltext{9}{Grove City High School, 4665 Hoover Road, Grove City, OH 43123, USA}
\altaffiltext{10}{Department of Physics, The Ohio State University, 191 W. Woodruff Ave., Columbs, OH 43210, USA}
\altaffiltext{11}{Warsaw University Astronomical Observatory, Al. Ujazdowskie 4, 00-478 Warsaw, Poland}
\altaffiltext{12}{Coral Towers Observatory, Cairns, Queensland 4870, Australia}
\altaffiltext{13}{Las Cumbres Observatory Global Telescope Network, 6740 Cortona Drive, Suite 102, Santa Barbara, CA 93117}
\footnotetext[$\dagger$]{This publication is partially based on observations obtained with the Apache Point Observatory 3.5-meter telescope, which is owned and operated by the Astrophysical Research Consortium.}

\begin{abstract}
We analyze a $\Delta V\sim-9$ magnitude flare on the newly identified M8 dwarf SDSS J022116.84+194020.4 (hereafter {\name}) detected as part of the All-Sky Automated Survey for Supernovae (ASAS-SN). Using infrared and optical spectra, we confirm that {\name} is a relatively nearby (d$\sim$76 pc) M8 dwarf with strong quiescent H$\alpha$ emission. Based on kinematics and the absence of features consistent with low-gravity (young) ultracool dwarfs, we place a lower limit of 200 Myr on the age of {\name}. When modeled with a simple, classical flare light-curve, this flare is consistent with a total $U$-band flare energy $E_U\sim$ 10$^{34}$ erg, confirming that the most dramatic flares are not limited to warmer, more massive stars. Scaled to include a rough estimate of the emission line contribution to the $V$ band, we estimate a blackbody filling factor of $\sim$10--30\% during the flare peak and $\sim$0.5--1.6\% during the flare decay phase. These filling factors correspond to flare areas that are an order of magnitude larger than those measured for most mid-M dwarf flares.
\end{abstract}


\section{Introduction}
\label{sec:intro}
M dwarfs are well known for their magnetic activity, both from quiescent H$\alpha$ emission \citep[e.g.,][]{Hawley1996} and dramatic flare events with emission spanning the entire electromagnetic spectrum \citep[e.g.,][]{Osten2005}. Though flares can be found across the entire M spectral class, flares are most often observed on mid-M dwarfs. Early-M dwarfs are on average less active than mid-M dwarfs, while late-M dwarfs are too faint for most flare-monitoring campaigns. Despite a low number of detected flares, quiescent activity is observed in a larger fraction of late-M dwarfs than mid-M dwarfs. In the Solar neighborhood, $\sim$80\% of M8 dwarfs show H$\alpha$ emission compared to 20\% of M3 dwarfs \citep{West2011}.  

The increase in the active fraction with spectral type is consistent with a changing relationship between activity, age, and rotation \citep{Reiners2010}. Active early-M dwarfs are found, on average, closer to the Galactic plane than active late-M dwarfs, indicating that late-M dwarfs are active for a longer portion of their lifetimes \citep{West2008}. Mid-M dwarfs that flare are found, on average, at lower Galactic heights than those with only H$\alpha$ emission \citep{Kowalski2009}, implying that the average flare lifetime is shorter than the quiescent activity lifetime. The age of flaring late-M dwarfs is particularly interesting because the M7-M9 spectral types include the most massive brown dwarfs at ages $<$1 Gyr \citep{Burrows1997}.

Flares on early- and mid-M dwarfs follow well characterized patterns; small flares ($E_U \sim 10^{28}$--$10^{30}$ ergs) typically occur hourly or daily, while larger flares ($E_U \sim 10^{32}$--$10^{34}$ ergs) typically occur no more often than weekly  \citep{Lacy1976}. These patterns vary with both spectral type and base activity level \citep{Hilton2011phd,Davenport2012}, but it is unclear whether late-M dwarfs flare more or less frequently than mid-M dwarfs. With $<$20 total detected late-M dwarf flares events \citep[e.g.,][]{Rockenfeller2006,Hilton2011phd,Berger2013}, the patterns followed by late-M dwarfs are unclear. 

On UT 2013 August 14, SDSS J022116.84+194020.4 (hereafter {\name}) was flagged as transient ASASSN-13cb in the All-Sky Automated Survey for Supernovae \citep[ASAS-SN;][]{Shappee2013} with a flare peak emission of $\Delta V\sim -9$. We present follow-up observations and use models based on mid-M dwarf flares to estimate the properties of the flare. In Section~\ref{sec:obs}, we present our observations and characterize {\name}, in Section~\ref{sec:flaremod} we examine the flare, and in Section~\ref{sec:disc} we place {\name} in the context of late-M dwarf magnetic activity. 

\section{Observations and Survey Data}
In addition to the detection of the flare, we examined the photometry available from sky surveys and obtained follow-up spectroscopy to investigate the properties of {\name}. Those data are described below. 

\label{sec:obs}
\subsection{Photometric Data}
\label{sec:surveys}
ASAS-SN\footnote{\url{http://www.astronomy.ohio-state.edu/$\scriptstyle\mathtt{\sim}$assassin/index.shtml}} is an optical transient survey that images the sky visible from Haleakala, Hawaii every $\sim 5$~days down to $V\sim17$, using two 14-cm telescopes in a common mount \citep[see]{Shappee2013}. The $V$-band images ($2\times 90$~sec exposures per field) are automatically processed through a difference imaging pipeline that produces transient candidates within $\sim 1$~hr of the initial observation. We discovered the bright $V\simeq 13$ transient ASASSN-13cb (R.\ A. = 02 21 16.92, Decl. = 19 40 19.9) on UT 2013 Aug 14.52 \citep{Stanek2013}. The transient faded by $\Delta V \sim 0.5$ between the two discovery images and by $\Delta V \sim 3.5$ in confirmation images obtained 2.3~hr later.  The photometry of ASASSN-13cb, presented in Table~1, was obtained from ASAS-SN images with aperture and PSF fitting photometry using IRAF and Daophot~II  \citep{Stetson1992} and calibrated using magnitudes of several stars from the AAVSO Photometric All-Sky Survey\footnote{\url{http://www.aavso.org/apass}}.

\begin{deluxetable}{lll} \tablewidth{0pt} \tabletypesize{\scriptsize}
\tablecaption{Flare Magnitudes and Fluxes \label{tab:lc} }
\tablehead{ \colhead{Time\tablenotemark{a}}  & \colhead{$V$} & \colhead{$F_V$} \\
 \colhead{(h:m:s)}  & \colhead{magnitude} & \colhead{(erg cm$^{-2}$ s$^{-1}$ \AA$^{-1}$ )} 
} 
\startdata
0:00:00  &  12.84 $\pm$ 0.03  &  $(2.68 \pm 0.07) \times 10^{-14}$  \\
0:01:57  &  13.33 $\pm$ 0.04  &  $(1.69 \pm 0.06) \times 10^{-14}$  \\
2:20:19  &  16.70 $\pm$ 0.14  & $ (7.65 \pm 0.96) \times 10^{-16}$  \\
\hline
        quiescent       & 22.09 $\pm$ 0.26  & (5.45 $\pm$ 1.52) $\times 10^{-18}$  
\enddata
\tablenotetext{a}{Since first flare detection (2013.621).}
\end{deluxetable}

We retrieved photometry from the Sloan Digital Sky Survey \citep[SDSS;][]{York2000}, the Two-Micron All Sky Survey \citep[2MASS;][]{Skrutskie2006}, and the Wilkinson Infrared Sky Explorer mission \citep[WISE;][]{Wright2010} based on a coordinate cross-match to the ASAS-SN position. Each survey returned only one source within our 5\arcsec~search radius. Poor quality flags were not set in any bands, but the uncertainties on the $u$, $W3$, and $W4$ measurements were sufficiently high to indicate unreliable magnitudes. The $grizJHK_SW1W2$ magnitudes are listed in Table~\ref{tab:prop}. 

\begin{deluxetable}{ll} \tablewidth{0pt} \tabletypesize{\scriptsize}
\tablecaption{Properties of {\name} in Quiescence \label{tab:prop} }
\tablehead{ \colhead{Parameter}  & \colhead{Value} } 
\startdata
\multicolumn{2}{c}{SDSS (2005.933)} \\
\hline
R.A. & \phs 02 21 16.84 \\
decl. & +19 40 20.4 \\
$g$ & \phs 22.80 $\pm$ 0.13 \\
$r$ &\phs  21.24 $\pm$ 0.05 \\
$i$ & \phs 18.65 $\pm$ 0.02 \\
$z$ & \phs 17.08 $\pm$ 0.02 \\
\hline
\multicolumn{2}{c}{2MASS (1997.805)} \\ 
\hline
R.A. & \phs 02 21 16.77   \\
decl. & +19 40 20.1 \\
$J$ &\phs  15.00 $\pm$ 0.04 \\
$H$ & \phs 14.44 $\pm$ 0.04 \\
$K_S$ &\phs  13.91 $\pm$ 0.05 \\
\hline
\multicolumn{2}{c}{WISE} \\ 
\hline
$W1$ &\phs  13.69 $\pm$ 0.03 \\
$W2$ & \phs 13.45 $\pm$ 0.04 \\
\hline
\multicolumn{2}{c}{Spectroscopic} \\ 
\hline
Spectral type & \phs M8 \\
H$\alpha$ EW & \phs 36 $\pm$ 11\AA \\
$L_{\rm H\alpha}/L_{\rm bol}$ & \phs  1.5$\times 10^{-4}$ \\
\hline
\multicolumn{2}{c}{Derived} \\ 
\hline
d ($M_r$ vs. $r-z$)  & \phs 78.36 $\pm$ 1.85 pc   \\
d ($M_r$ vs. $r-i$) & \phs 91.81 $\pm$ 2.27 pc  \\
d ($M_r$ vs.  $i-z$) & \phs 58.56 $\pm$ 1.63 pc  \\
d (mean) & \phs 76 $\pm$ 6 pc \\
$\mu_{\alpha}$ & $-$96.0 $\pm$ 6.7 mas yr$^{-1}$ \\
$\mu_{\delta}$ & $-$36.3 $\pm$ 7.9 mas yr$^{-1}$ \\
$V_{\rm tan}$ & \phs 37 $\pm$ 8 km s$^{-1}$ 
\enddata
\end{deluxetable}

The colors of {\name} are consistent with the median color of M8 dwarfs in the $g$ through $W2$ bands. It is not peculiar in its $g-r$ color \citep[used to select metal-poor M dwarfs;][]{Lepine2008} or its $J-K_S$ color \citep[consistent with young ultracool dwarfs;][]{Cruz2009}.  Distances calculated using the \citet{Bochanski2010} color-magnitude relations are given in Table~\ref{tab:prop}, with uncertainties including both magnitude uncertainties and the scatter in the relations. We adopt a mean distance of $d = 76 \pm 6$ pc; the uncertainty is dominated by the dispersion of the three distance estimates. 

We measured a proper motions of $\mu_{\alpha}=-96.0\pm6.7$~mas yr$^{-1}$ and $\mu_{\delta}=-36.3\pm7.9$~mas yr$^{-1}$ based on the difference between SDSS and 2MASS coordinates\footnote{The astrometric calibration of SDSS and 2MASS shows very good agreement ($<$0\farcs06) within typical coordinate uncertainties \citep[$\sim0\farcs1$;][]{Pier2003}.}. The combination of the distance and proper motion results in a tangential velocity of 37 $\pm$ 8 km s$^{-1}$, placing it slightly faster than the median V$_{\rm tan}$ for ultracool dwarfs near the Sun \citep[e.g.,][]{Faherty2009}. Based on the Bayesian statistical proper motion models of \citet{Malo2013}\footnote{Using the web-based tool at \url{http://www.astro.umontreal.ca/$\scriptstyle\mathtt{\sim}$malo/banyan.php}}, the kinematics of {\name} are not consistent with any of the seven closest and youngest moving groups, implying an age $>$100 Myr.

\subsection{Spectroscopic Data}
\begin{figure}
\includegraphics[width=0.95\linewidth]{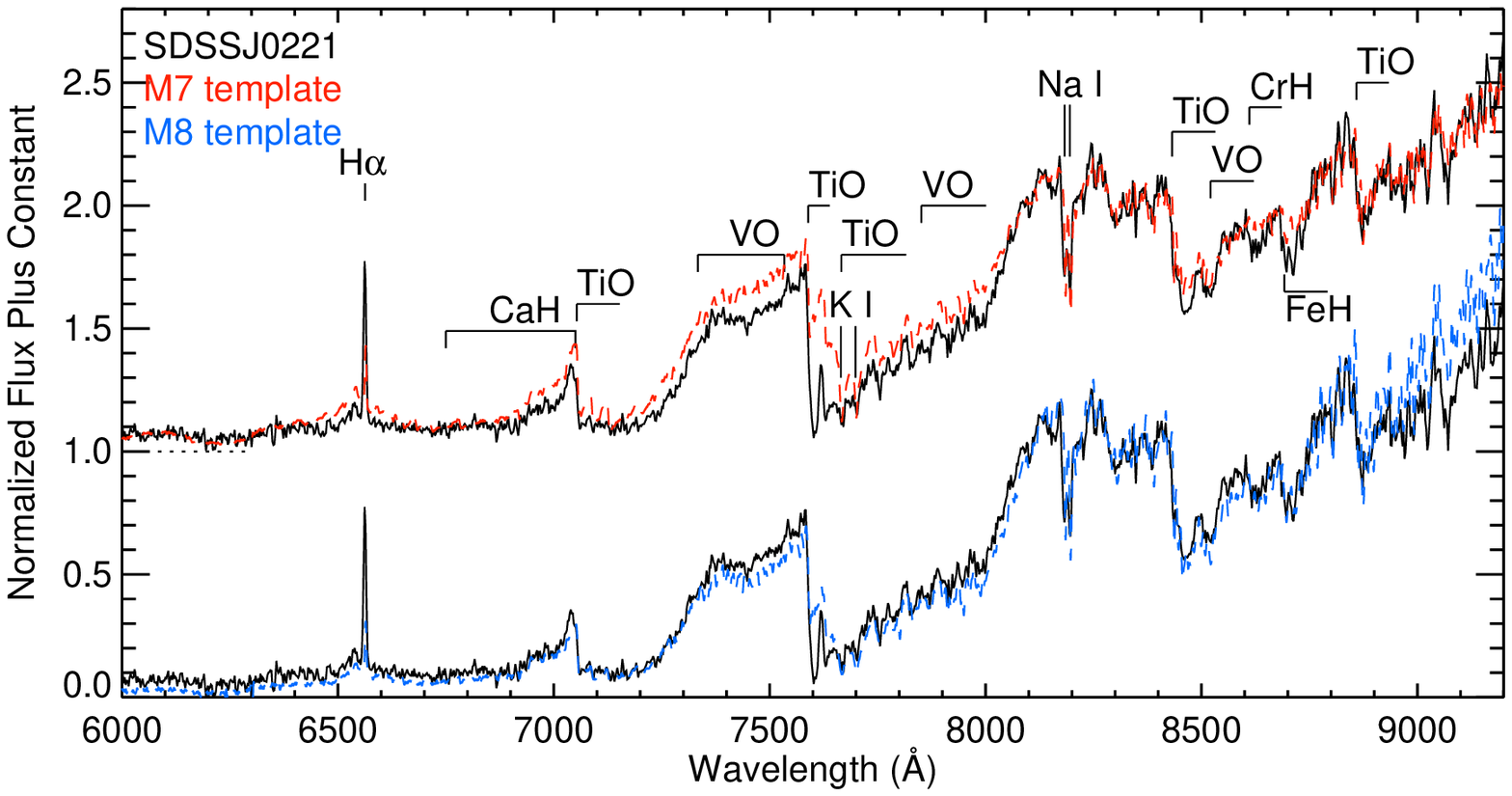}
\includegraphics[width=0.95\linewidth]{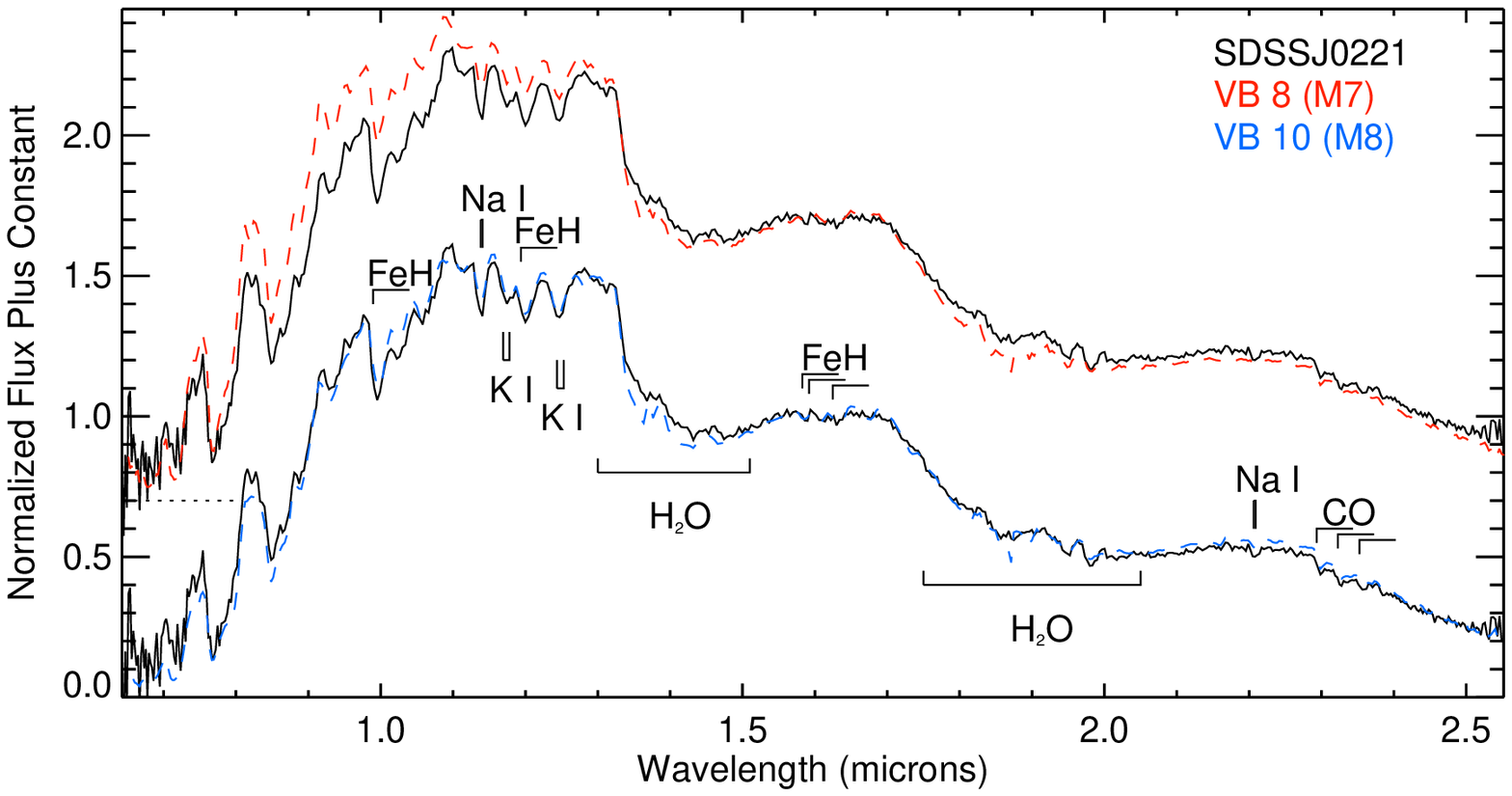} 
\caption{The quiescent optical (top panel) and infrared (bottom panel) spectra of {\name}. The spectra are normalized at 8350\AA~and 1.6 microns respectively and offset by constants for clarity (dotted lines). The optical spectrum is compared to the M7 (red) and M8 (blue) template spectra from \citet{Bochanski2007a} and the infrared spectrum is compared to the M7 dwarf VB 8 (red) and the M8 dwarf VB 10 (blue) from \citet{Burgasser2004}. Important atomic and molecular features are labeled.} \label{fig:newsp}
\end{figure}

We obtained low-resolution ($R\sim 800$) optical spectra of {\name} on three different nights (UT 2013 Aug 30 and Sep 1--2) using the Dual Imaging Spectrograph (DIS) on the ARC 3.5m telescope at APO and the Wide Field CCD Camera and Spectrograph (WFCCD) on the du Pont 2.5m telescope at LCO. We used the B400/R300 gratings and a 1\farcs5 slit (3500-10000\AA) with DIS and the 400~l/mm grism and a 1\farcs7 slit with WFCCD (3700-9500\AA). The spectra were reduced using LAcosmic  \citep{van-Dokkum2001} for cosmic ray rejection and standard techniques in the IRAF {\tt twodspec} and {\tt onedspec} packages for spectral extraction and wavelength+flux calibration. The median combined spectrum is shown in Figure~\ref{fig:newsp}.

From these spectra, we obtain a spectral type of M8 for {\name} using the automatic Hammer routines \citep[based on spectral indices;][]{West2004,Covey2007}. The optical spectrum is shown compared to the M7 and M8 templates from \citet{Bochanski2007a} in Figure~\ref{fig:newsp}. Visual comparison shows that the spectral features of {\name} place it between the two types, but the spectrum is a better match to the M8 template, indicating an M8 type. We were unable to measure radial velocities from the individual spectra due to their low resolution. 

We measure the H$\alpha$ equivalent width (EW) of the individual spectra using the Hammer \citep{West2004,Covey2007}, obtaining a mean and standard deviation of $36 \pm 11$\AA~(listed in Table~\ref{tab:prop}) for seven total spectra. Using a $\chi$ value from \citet[][in prep.]{Schmidt2013a}, we calculate an activity strength of $L_{\rm H\alpha}/L_{\rm bol} = 1.5\times 10^{-4}$. This is well above the median and dispersion for H$\alpha$ emission from M8 dwarfs \citep{West2011}. The full range of the measured H$\alpha$ EWs (22--54 \AA; $(0.9$--$2.3) \times 10^{-4}$) is larger compared to its mean H$\alpha$ EW than the range seen for an average late-M dwarf \citep{Bell2012}, but not remarkable given the strong emission and observed flare. 

We obtained a low-resolution (R$\sim$150) infrared spectrum on 2013 Sept 3 using SpeX \citep{Rayner2003} on the Infrared Telescope Facility (IRTF), as well as standard calibration frames and a spectrum of the A0 star HD 16811. The data were reduced and telluric-corrected using SpexTool \citep{Vacca2003,Cushing2004}. The H$_2$O indices compiled by \citet{Allers2013} indicate an infrared spectral type of M7, but direct comparison to spectral standards (shown in Figure~\ref{fig:newsp}) results in a final infrared spectral type of M8. 

The infrared spectrum also includes features that are sensitive to the surface gravity (age) of ultracool dwarfs. We calculated the the FeH, VO, K I, and H-cont indices from \citet{Allers2013}; the combined score from all four indices indicates features consistent with a typical field dwarf, placing a rough lower limit of 200 Myr on the age of {\name}. While the Li I absorption line could provide an additional limit on the age of {\name}, the optical spectrum does not have sufficient S/N and resolution to place limits on Li I.

\section{Properties of the Flare}
\label{sec:flaremod}
\begin{figure}
\includegraphics[width=0.95\linewidth]{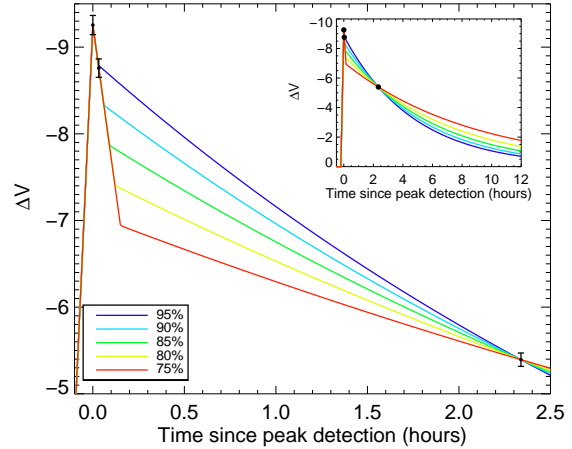} 
\caption{$\Delta V$ as a function of time. The three ASAS-SN detections and associated uncertainties are shown (black points) with five different lightcurves generated assuming a simple, classical flare model (colored lines). The flare model assumes an impulsive phase lasting until the magnitude of the flare decreases to a percentage of the observed peak magnitude. It is likely that the actual flare morphology was more complex, either with minor variations about the classic shape or major substructure.} \label{fig:lc}
\end{figure}

We calculated a quiescent $V$-band magnitude of $V$ = 22.09 $\pm$ 0.26 for {\name} by calibrating the M8 template spectrum to the SDSS $r$-band magnitude and then integrating the spectrum over a $V$-band filter curve. This quiescent magnitude results in $\Delta V  = - 9.25 \pm 0.26$ for the observed peak magnitude (shown in Figure~\ref{fig:lc}). It is unlikely that we observed the flare at its true peak. However, more energetic flares occur are increasinly infrequent \citep{Lacy1976}, so it is also unlikely that the flare was much larger than observed. \citet{Kowalski2013} includes flares with impulsive decays lasting from 0.02 to 0.2 hours. If we assume a linear impulsive decay, this range of decay times is consistent with a range of peak magnitudes from the observed $\Delta V \sim -9.25$ (at $t=0$~hr) to a $\Delta V \sim -12$ (at $t=-0.18$~hr) peak. For simplicity, we assume that the observed peak is also the total flare peak. 

\subsection{The Flare Lightcurve}
Classical flares have a characteristic shape: an ``impulsive" phase that includes a fast rise and decline (well approximated as linear changes in magnitude with time) and a ``gradual" phase typically modeled by an exponential decay. Flare lightcurves have a wide variety of morphologies beyond simple, classical flares; some show multiple peaks \citep[e.g.,][]{Hawley1991} or complex structures in their decay phase \citep[e.g.][]{Kowalski2010}. With only three detections of {\name} during the flare, we cannot place constraints on the flare morphology, but we can use a classical flare model to estimate the total energy of the flare. 

We model the rise phase as a linear rise in $\Delta V$ with time, estimating a rise time of 0.2 hours \citep[reasonable for a large flare; e.g.,][]{Hawley1991,Kowalski2010}. The first two detections indicate a strong impulsive decay, lasting at minimum the two minutes between the first two observations. The third detection is consistent with a long gradual decay after a short impulsive phase. We model the impulsive decay as a linear fit to the first two points with a minimum length set to a 95\% decay from the peak and a maximum length of a 75\% decay from the peak. The maximum length is limited by the need to end at a magnitude bright enough so that the exponential decay does not fall below $\Delta V = -5.4$ at $t=2.3$~hr. The gradual phase is modeled by an exponential decay beginning after the impulsive phase with a timescale consistent with the gradual phase detection.

We can estimate the equivalent duration (ED; the time required to emit the flare flux during quiescence) and total energy of the flare by integrating over the range of model lightcurves. Adopting a quiescent flux of $F_V = (5.5 \pm 1.5) \times 10^{-18}$ ergs cm$^{-2}$ s$^{-1}$ \AA$^{-1}$, we measure EDs of $(5.2$--$9.0) \times 10^6$~s for the five models. These are much larger than the $U$-band EDs for typical flares \citep[$\sim$1000~s]{Kowalski2013,Walkowicz2011} and one order of magnitude larger than the $U$-band ED for the largest flares detected \citep{Kowalski2010}

Using a quiescent luminosity of $L_V = (3.4 \pm 0.9) \times 10^{27} $ ergs s$^{-1}$, we calculate a total $V$-band flare energy of $E_V = (1.8$--$3.7) \times 10^{34}$ ergs. The conversion from $E_V$ to $E_U$ from \citet{Lacy1976} indicates that the $U$-band energy of this flare is $E_U= (3.2$--$5.5) \times 10^{34}$ ergs. The only flares with comparable published $U$-band energies are the giant flares on AD Leo \citep{Hawley1991} and YZ CMi \citep{Kowalski2010}, both at just over $E_U = 10^{34}$ ergs. While the $E_U$ calculated for the flare on {\name} is slightly larger, its uncertainties due to the estimation of the flare shape and the use of scaling relations are likely to be an order of magnitude. Extrapolating the flare frequency distributions of \citet{Hilton2011phd} to $E_U = 10^{34}$ ergs, flares this large should occur on active mid-M dwarfs monthly, and active late-M dwarfs once per year.

\subsection{Emission in the $V$-band}
At optical and UV wavelengths, flare emission originates from two components. The major contributor is a $T\sim10000$~K blackbody \citep[e.g.,][]{Hawley1992} thought to originate deep in the stellar atmosphere near the foot points of the magnetic field loops. Atomic emission lines \citep[e.g.,][]{Fuhrmeister2010} and hydrogen Balmer continuum \citep{Kunkel1970} are emitted as part of a second, lower density component. The continuum emission dominates the overall optical/UV energy budget of the flare, contributing 91--95\% of the total  during the impulsive phase and 69--95\% during the gradual phase \citep{Hawley1991}. 

Spectroscopic observations of the $V$ band during flares are rare, in part because there are only two strong emission lines, H$\beta$ and He I $\lambda$5876. \citet{Hawley2003} calculated an energy budget for four flares on AD Leo with spectra overlapping the $V$-band filter. They found that the continuum contributes 89--96\% of the $V$-band energy budget during the impulsive phase and 0--95\% of the $V$-band energy budget during the gradual phase. The large range of continuum emission in the $V$ band during the quiescent phase is due to the faintness of the blackbody compared to the line emission and stellar flux; in a large flare, the blackbody is likely to remain strong even during the decay phase. 

We can examine the range of filling factors and blackbody emission temperatures consistent with the impulsive and gradual phase observations by calculating the blackbody contribution to the flux, $F_{\lambda}$, as
\begin{equation}
F_{\lambda} = X \frac{R^2}{d^2} \pi B_{\lambda} (T),
\end{equation}
where $X$ is the filling factor of the blackbody spectrum, $R$ is the stellar radius, $d$ is the distance, and $T$ is the characteristic temperature of the blackbody distribution. We adopt $R = 0.124$~R$_{\odot}$ \citep[the radius derived for M8 LP 349-25B;][]{Dupuy2010} and $d=76$~pc (Section~\ref{sec:surveys}) for the radius and distance. 

\begin{figure}
\includegraphics[width=0.95\linewidth]{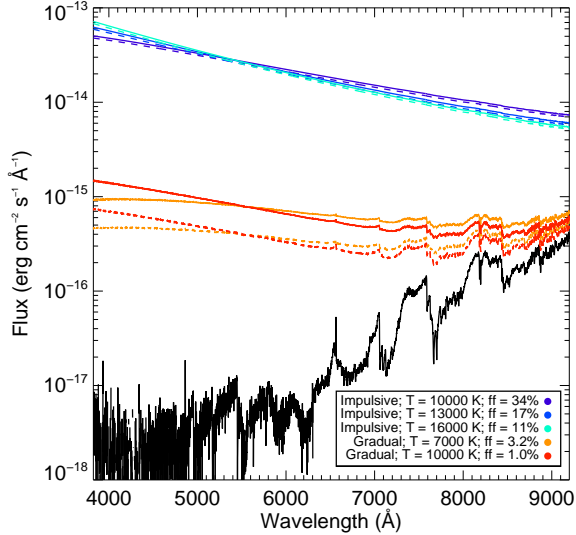} 
\caption{The M8 template spectrum calibrated to the SDSS magnitudes of {\name} (in quiescence; black) with blackbody emission curves shown to simulate the peak (green, blue, and purple lines) and gradual phase (red and orange lines) magnitudes. The solid lines show filling factors of 100\% of the $V$-band flare emission originating in the blackbody continuum, while the dashed lines show filling factors of 95\% (impulsive phase) and 50\% (gradual phase) of the $V$-band emission originating in the blackbody continuum.} \label{fig:flarespec}
\end{figure}

\citet{Kowalski2013} directly fit blackbody functions to blue optical spectra of flaring mid-M dwarfs, obtaining temperatures from $T = 9800$ to $14100$~K for the peak and $T = 5600$ to $8900$~K during the decay phase of impulsive flares. As the flare on {\name} was larger than most of the flares examined, we adopt the slightly higher values to examine the area coverage of continuum emission; $T = 10000$, $13000$, and $16000$~K for the impulsive phase and $T=7000$ and $10000$~K for the gradual phase. The resulting model spectra are shown in Figure~\ref{fig:flarespec}, both with the blackbody modeled as the only contribution to the flare $V$-band flux and scaled so that the blackbody contributes 95\% during the impulsive phase and 50\% during the gradual phase. 

During the impulsive phase, blackbody emission with a characteristic temperature of $T = 10000$, $13000$, and $16000$~K would need to have filling factors of 32\%, 16\%, and 11\% respectively to produce 95\% of the observed flare emission. Those filling factors correspond to a physical area of (2--8) $\times 10^{19}$ cm$^2$ on an object this size, while the largest of the mid-M dwarf flares with detected peaks \citep{Hawley2003,Fuhrmeister2008,Kowalski2013} have flare surface areas of $(1$--$4) \times 10^{18}$ cm$^2$ (0.05--0.2\% of a R = 0.35R$_{\odot}$ star). During the gradual phase, a $T=10000$ or $7000$~K blackbody spectrum would need filling factors of 0.5\% or 1.6\% respectively to produce 50\% of the decay phase emission. These filling factors are similar to the physical area covered by the giant flare from \citet{Kowalski2010} during its decay phase.

\section{Discussion}
\label{sec:disc}
The low-gravity and kinematic indicators often used to select young brown dwarfs among ultracool dwarf populations indicate that {\name} is older than 200 Myr. According to ultracool dwarf evolutionary models \citep{Burrows1997} an M8 dwarf (corresponding to a $\sim2800$~K surface temperature) of $<$1 Gyr could be a massive brown dwarf, but with thin disk kinematics, {\name} is more likely a few Gyr old and so is probably a star rather than a brown dwarf. Stars with dramatic flares are typically assumed to be young, but with the break down of the age-activity relation \citep{Reiners2010} and the persistence of quiescent activity in late-M dwarfs for $\sim$8 Gyr \citep{West2008}, it is possible that ultracool dwarfs can have large flares even at typical thin disk ages. 

The $E_U \sim 10^{34}$ erg estimate of the energy released is comparable to the highest energies calculated for mid-M dwarf flares \citep{Hawley1991,Kowalski2010}. The flare is likely to have covered $>$10\% of the stellar (or possibly brown dwarf) surface at its peak magnitude, significantly larger than the area coverage at the peaks of most flares \citep{Hawley2003,Fuhrmeister2008,Kowalski2013}, but comparable to that of the largest flares \citep{Hawley1991,Kowalski2010}.

The flare on {\name} is not the only very large amplitude flare detected on a late-M dwarf; \citet{Schaefer1990} report a very similar flare on CZ Cnc, and strong flares have been observed at other wavelengths \citep[e.g.,][]{Tagliaferri1990,Fleming2000} and through optical spectroscopy \citep[e.g.,][]{Liebert1999,Schmidt2007}. Overall, however, there are not yet sufficient observations to characterize the flare frequency distribution of M7-M9 dwarfs and investigate the similarity of their emission mechanisms to those on more massive M dwarfs. 

\acknowledgements
The authors thank A.\ F.\ Kowalski for useful suggestions that improved this manuscript.

 J.\ F.\ B.\ is supported by NSF grant PHY-1101216. Development of ASAS-SN has been supported by NSF grant AST-0908816 and the Center for Cosmology and AstroParticle Physics at The Ohio State University. 

This research has benefitted from the SpeX Prism Spectral Libraries, maintained by Adam Burgasser at \url{http://pono.ucsd.edu/$\sim$adam/browndwarfs/spexprism} and through the use of the AAVSO Photometric All-Sky Survey (APASS), funded by the Robert Martin Ayers Sciences Fund. This research has also made use of NASA's Astrophysics Data System.

This publication makes use of data products from the Two Micron All Sky Survey, which is a joint project of the University of Massachusetts and the Infrared Processing and Analysis Center/California Institute of Technology, funded by the National Aeronautics and Space Administration and the National Science Foundation. This publication also makes use of data products from the Wide-field Infrared Survey Explorer, which is a joint project of the University of California, Los Angeles, and the Jet Propulsion Laboratory/California Institute of Technology, funded by the National Aeronautics and Space Administration.

Funding for SDSS-III has been provided by the Alfred P. Sloan Foundation, the Participating Institutions, the National Science Foundation, and the U.S. Department of Energy Office of Science. The SDSS-III web site is \url{http://www.sdss3.org/}. SDSS-III is managed by the Astrophysical Research Consortium for the Participating Institutions (listed at \url{http://www.sdss3.org/collaboration/boiler-plate.php}).


\end{document}